# A Population of Fast Radio Bursts at Cosmological Distances

D. Thornton,[1,2*] B. Stappers,[1] M. Bailes,[3,4] B. Barsdell,[3,4] S. Bates,[5] N. D. R. Bhat,[3,4,6] M. Burgay,[7] S. Burke-Spolaor,[8] D. J. Champion,[9] P. Coster,[2,3] N. D'Amico,[10,7] A. Jameson,[3,4] S. Johnston,[2] M. Keith,[2] M. Kramer,[9,1] L. Levin,[5] S. Milia,[7] C. Ng,[9] A. Possenti,[7] W. van Straten[3,4]

[1] Jodrell Bank Centre for Astrophysics, School of Physics and Astronomy, University of Manchester, Manchester, M13 9PL, UK.

[2] Commonwealth Science and Industrial Research Organisation (CSIRO) Astronomy and Space Science, Australia Telescope National Facility, Post Office Box 76, Epping, NSW 1710, Australia.

[3] Centre for Astrophysics and Supercomputing, Swinburne University of Technology, Mail H30, Post Office Box 218, Hawthorn, VIC 3122, Australia.

[4] Australian Research Council Centre of Excellence for All-Sky Astrophysics (CAASTRO), Mail H30, Post Office Box 218, Hawthorn, VIC 3122, Australia.

[5] West Virginia University Center for Astrophysics, West Virginia University, Morgantown, WV 26506, USA.

[6] International Centre for Radio Astronomy Research, Department of Imaging and Applied Physics, Faculty of Science and Engineering, Curtin University, Post Office Box U1987, Perth, WA 6845, Australia.

[7] Instituto Nazionale di Astrofisica, Osservatorio Astronomico di Cagliari, Loc. Poggio dei Pini, Strada 54, 09012 Capoterra (CA), Italy.

[8] Jet Propulsion Laboratory, California Institute of Technology, 4800 Oak Grove Drive, Pasadena, CA 91104, USA.

[9] Max-Planck-Institut für Radio Astronomie, Auf dem Hugel 69, 53121 Bonn, Germany.

[10] Dipartimento di Fisica, Universita di Cagliari, Cittadella Universitaria 09042, Monserrato (CA), Italy.

*Corresponding author. E-mail: thornton@jb.man.ac.uk

**Searches for transient astrophysical sources often reveal unexpected classes of objects that are useful physical laboratories. In a recent survey for pulsars and fast transients we have uncovered four millisecond-duration radio transients all more than 40° from the Galactic plane. The bursts' properties indicate that they are of celestial rather than terrestrial origin. Host galaxy and intergalactic medium models suggest that they have cosmological redshifts of 0.5 to 1, and distances of up to 3 gigaparsecs. No**

**temporally coincident x- or gamma-ray signature was identified in association with the bursts. Characterization of the source population and identification of host galaxies offers an opportunity to determine the baryonic content of the Universe.**

**Main Text:**

The four Fast Radio Bursts, FRBs, (Fig. 1) reported here were detected in the high Galactic latitude region of the High Time Resolution Universe (HTRU) survey (*1*) which was designed to detect short-time-scale radio transients and pulsars (Galactic pulsed radio sources). The survey uses the 64-m Parkes radio telescope and its 13-beam receiver to acquire data across a bandwidth of 400 MHz centered at 1.382 GHz (Table S1). We measure minimum fluences for the FRBs of $F$ = 0.6 to 8.0 Jy ms (*2*). At cosmological distances this indicates they are more luminous than bursts from any known transient radio source (*3*). Follow-up observations at the original beam positions have not detected any repeat events indicating that the FRBs are likely cataclysmic in nature.

Candidate extragalactic bursts have previously been reported with varying degrees of plausibility (*4–7*) along with a suggestion that FRB 010724 (the "Lorimer burst") is similar to other signals that may be of local origin (*8, 9*). To be consistent with a celestial origin, FRBs should exhibit certain pulse properties. In particular, observations of radio pulsars in the Milky Way (MW) have confirmed that radio emission is delayed by propagation through the ionized interstellar medium (ISM) which can be considered a cold plasma. This delay has a power law dependence of $\delta t \propto \mathrm{DM} \cdot \nu^{-2}$ and typical frequency dependent width of $W \propto \nu^{-4}$. The dispersion measure (DM) is related to the integrated column density of free electrons along the line of sight to the source and is a proxy for distance. The frequency-dependent pulse broadening occurs as an astrophysical pulse is scattered by an inhomogeneous turbulent medium, causing a characteristic exponential tail. Parameterizing the frequency dependence of $\delta t$ and $W$ as $\alpha$ and $\beta$ respectively we measure $\alpha = -2.003 \pm 0.006$ and $\beta = -4.0 \pm 0.4$ for FRB 110220 (Table 1 and Fig. 2), as expected for propagation through a cold plasma. Although FRB 110703 shows no evidence of scattering, we determine $\alpha = -2.000 \pm 0.006$. The other FRBs do not have sufficient signal-to-noise ratios (SNRs) to yield astrophysically interesting constraints for either parameter and show no evidence of scattering.

Our FRBs were detected with DMs in the range 553 to 1103 $cm^{-3}$ pc. Their high Galactic latitudes ($|b| > 41°$, Table 1) correspond to lines of sight through the low column density Galactic ISM corresponding to just 3 to 6% of the DM measured (*10*). These small Galactic DM contributions are highly supportive of an extragalactic origin and are substantially smaller fractions than those of previously reported bursts, which were 15% of DM = 375 $cm^{-3}$ pc for FRB 010724 (*4*) and 70% of DM = 746 $cm^{-3}$ pc for FRB 010621 (*5*).

The non-Galactic DM contribution, $DM_E$, is the sum of two components: the intergalactic medium (IGM; $DM_{IGM}$), and a possible host galaxy ($DM_{Host}$). The intervening medium could be purely intergalactic and could also include a contribution from an intervening galaxy. Two options are considered according to the proximity of the source to the center of a host galaxy.

If located at the center of a galaxy, this may be a highly-dispersive region; for example lines of sight passing through the central regions of Milky Way-like galaxies could lead to DMs in excess of 700 cm$^{-3}$ pc in the central ~100 pc (*11*), independent of the line-of-sight inclination. In this case $DM_E$ is dominated by $DM_{Host}$ and requires FRBs to be emitted by an unknown mechanism in the central region, possibly associated with the supermassive black hole located there.

If outside a central region, then elliptical host galaxies (which are expected to have a low electron density) will not contribute to $DM_E$ substantially, and $DM_{Host}$ for a spiral galaxy will only contribute substantially to $DM_E$ if viewed close to edge-on (inclination, $i > 87°$ for $DM > 700$ cm$^{-3}$ pc ; probability $(i > 87°) \approx 0.05$). The chance of all four FRBs coming from edge-on spiral galaxies is therefore negligible (10$^{-6}$). Consequently, if the sources are not located in a galactic center, $DM_{Host}$ would likely be small, and $DM_{IGM}$ dominates. Assuming an IGM free electron distribution, which takes into account cosmological redshift and assumes a Universal ionization fraction of 1 (*12, 13*), the sources are inferred to be at redshifts $z = 0.45$ to $0.96$, corresponding to co-moving distances of 1.7 to 3.2 Gpc (Table 1).

In principle, pulse scatter-broadening measurements can constrain the location and strength of an intervening scattering screen (*14*). FRBs 110627, 110703, and 120127 are too weak to enable the determination of any scattering; however, FRB 110220 exhibits an exponential scattering tail (Fig. 1). There are at least two possible sources and locations for the responsible scattering screens: a host galaxy or the IGM. It is possible that both contribute to varying degrees.

For a screen-source, $D_{src}$, and screen-observer, $D_{obs}$, distances, the magnitude of the pulse broadening due to scattering is multiplied by the factor $D_{src} D_{obs} / (D_{src} + D_{obs})^2$. For a screen and source located in a distant galaxy this effect probably requires the source to be in a high-scattering region, for example, a galactic center.

The second possibility is scattering because of turbulence in the ionized IGM, unassociated with any galaxy. There is a weakly constrained empirical relationship between DM and measured scattering for pulsars in the MW. If applicable to the IGM, then the observed scattering implies, $DM_{IGM} > 100$ cm$^{-3}$ pc (*2, 15*). With use of the aforementioned model of the ionized IGM, this DM equates to $z > 0.11$ (*2, 12, 13*). The probability of an intervening galaxy located along the line of sight within $z \approx 1$ is $\leq 0.05$ (*16*). Such a galaxy could be a source of scattering and dispersion, but the magnitude would be subject to the same inclination dependence as described for a source located in the disk of a spiral galaxy.

It is important to be sure that FRBs are not a terrestrial source of interference. Observations at Parkes have previously shown swept frequency pulses of terrestrial origin, dubbed "perytons". These are symmetric $W > 20$ ms pulses which imperfectly mimic a dispersive sweep (*2, 8*). Although perytons peak in apparent DM near 375 cm$^{-3}$ pc (range from ~200 to 420 cm$^{-3}$ pc), close to that of FRB 010724, the FRBs presented here have much higher and randomly-distributed DMs. Three of these FRBs are factors of > 3 narrower than any documented peryton. Last, the characteristic scattering shape and strong dispersion delay adherence of FRB 110220 make a case for cold plasma propagation.

The Sun is known to emit frequency swept radio bursts at 1–3 GHz (typeIIIdm; *17*). These bursts have typical widths of 0.2 to 10 seconds and positive frequency sweeps, entirely inconsistent with measurements of *W* and *α* for the FRBs. Whilst FRB 110220 was separated from the Sun by 5.6°, FRB 110703 was detected at night and the others so far from the Sun that any solar radiation should have appeared in multiple beams. These FRBs were only detected in a single beam; it is therefore unlikely they are of solar origin.

Uncertainty in the true position of the FRBs within the frequency-dependent gain pattern of the telescope makes inferring a spectral index, and hence flux densities outside the observing band, difficult. A likely off-axis position changes the intrinsic spectral index substantially. The spectral energy distribution across the band in FRB 110220 is characterized by bright bands ~100 MHz wide (Fig. 2); the SNRs are too low in the other three FRBs to quantify this behavior (*2*). Similar spectral characteristics are commonly observed in the emission of high-|*b*| pulsars.

With four FRBs, it is possible to calculate an approximate event rate. The high-latitude HTRU survey region is 24% complete, resulting in 4500 square degrees observed for 270 seconds. This corresponds to an FRB rate of
$R_{FRB}(F \sim 3 \text{ Jy ms}) = 1.0^{+0.6}_{-0.5} \times 10^4$ sky$^{-1}$ day$^{-1}$ where the 1-σ uncertainty assumes Poissonian statistics. The MW foreground would reduce this rate, with increased sky temperature, scattering, and dispersion for surveys close to the Galactic plane. In the absence of these conditions our rate implies that $17^{+9}_{-7}$, $7^{+4}_{-3}$ and $12^{+6}_{-5}$ FRBs should be found in the completed high- and medium-latitude parts of the HTRU (*1*) and Parkes multibeam pulsar (PMPS) surveys (*18*).

One candidate FRB with $DM > DM_{MW}$ has been detected in the PMPS (|*b*| < 5°; *5,19*), but this burst could be explained by neutron star emission, given a small scale-height error; however observations have not detected any repetition. No excess-DM FRBs were detected in a burst search of the first 23% of the medium-latitude HTRU survey (|*b*| < 15°; *20*).

The event rate originally suggested for FRB 010724, $R_{010724} = 225$ sky$^{-1}$ day$^{-1}$ (*4*), is consistent with our event rate, given a Euclidean Universe and a population with distance-independent intrinsic luminosities (source count, $N \propto F^{-3/2}$) giving
$R_{FRB}(F \sim 3 \text{ Jy ms}) \sim 10^2 R_{FRB}(F_{010724} \sim 150 \text{ Jy ms})$.

There are no known transients detected at gamma-ray, x-ray or optical wavelengths or gravitational wave triggers that can be temporally associated with any FRBs. In particular there is no known gamma-ray burst (GRB) with a coincident position on a timescale commensurate with previous tentative detections of short-duration radio emission (*6*). GRBs have highly beamed gamma-ray emission (*21*) and, if FRBs are associated with them, the radio emission must be beamed differently. Using the distances in Table 1 the co-moving volume contains ~$10^9$ late-type galaxies (*22*) and the FRB rate is therefore $R_{FRB} \sim 10^{-3}$ per galaxy per year . $R_{FRB}$ is thus inconsistent with $R_{GRB} \sim 10^{-6}$ per galaxy per year , even when beaming of emission is accounted for (*21*). Soft gamma-ray repeaters (SGRs) undergo giant bursts at a rate consistent with FRBs (*23*) and the energy within our band is well within the budget of the few known SGR giant burst cases (*24*).

Another postulated source class is the interaction of the magnetic fields of two coalescing neutron stars (*25*). However, the large implied FRB luminosities indicate that coalescing neutron stars may not be responsible for FRBs. Furthermore, $R_{FRB}$ is substantially higher than the predicted rate for neutron star mergers. Black hole evaporation has also been postulated as a source of FRBs; however the predicted luminosity within our observing band far exceeds the energy budget of an evaporation event (*26*).

The core-collapse supernova (ccSN) rate of $R_{ccSN} \sim 10^{-2}$ per galaxy per year (*27*) is consistent with $R_{FRB}$. There is no known mechanism to generate an FRB from a lone ccSN. It may, however, be possible that a ccSN with an orbiting neutron star can produce millisecond-duration radio bursts during the interaction of the ccSN explosion and the magnetic field of the neutron star (*28*), although the need for an orbiting neutron star will make these rarer.

As extragalactic sources, FRBs represent a probe of the ionized IGM. Real-time detections and immediate follow-up at other wavelengths may identify a host galaxy with an independent redshift measurement, thus enabling the IGM baryon content to be determined (*12*). Even without host identifications, further bright FRB detections will be a unique probe of the magneto-ionic properties of the IGM.

of data obtained from the High Energy Astrophysics Science Archive Research Center, provided by NASA's Goddard Space Flight Center. Part of this research was carried out at the Jet Propulsion Laboratory, California Institute of Technology, under a contract with NASA. The Parkes radio telescope is part of the Australia Telescope National Facility, which is funded by the Commonwealth of Australia for operation as a National Facility managed by CSIRO. Part of this research was conducted because of the support of CAASTRO through project number CE110001020. D.T. gratefully acknowledges the support of the Science and Technology Facilities Council and the CSIRO Astronomy and Space Science in his Ph.D. studentship. N.D.R.B. is supported by a Curtin Research Fellowship (CRF12228).

**Supplementary Materials**
www.sciencemag.org
Materials and Methods
Figs. S1 to S4
Table S1
References (*30*, *31*)

| Name | FRB 110220 | FRB 110627 | FRB 110703 | FRB 120127 |
|---|---|---|---|---|
| Beam Right Ascension (J2000) | $22^h\ 34^m$ | $21^h\ 03^m$ | $23^h\ 30^m$ | $23^h\ 15^m$ |
| Beam Declination (J2000) | $-12°\ 24'$ | $-44°\ 44'$ | $-02°\ 52'$ | $-18°\ 25'$ |
| Galactic Latitude, $b$ (°) | −54.7 | −41.7 | −59.0 | −66.2 |
| Galactic Longitude, $l$ (°) | +50.8 | +355.8 | +81.0 | +49.2 |
| UTC (dd/mm/yyyy hh:mm:ss.sss) | 20/02/2011 01:55:48.957 | 27/06/2011 21:33:17.474 | 03/07/2011 18:59:40.591 | 27/01/2012 08:11:21.723 |
| DM ($cm^{-3}$ pc) | 944.38 ± 0.05 | 723.0 ± 0.3 | 1103.6 ± 0.7 | 553.3 ± 0.3 |
| $DM_E$ ($cm^{-3}$ pc) | 910 | 677 | 1072 | 521 |
| Redshift, $z$ ($DM_{Host}$ = 100 $cm^{-3}$ pc) | 0.81 | 0.61 | 0.96 | 0.45 |
| Co-moving Distance, $D$ (Gpc) at $z$ | 2.8 | 2.2 | 3.2 | 1.7 |
| Dispersion Index, $\alpha$ | −2.003 ± 0.006 | – | −2.000 ± 0.006 | – |
| Scattering Index, $\beta$ | −4.0 ± 0.4 | – | – | – |
| Observed Width at 1.3 GHz, $W$ (ms) | 5.6 ± 0.1 | < 1.4 | < 4.3 | < 1.1 |
| SNR | 49 | 11 | 16 | 11 |
| Minimum Peak Flux Density $S_\nu$ (Jy) | 1.3 | 0.4 | 0.5 | 0.5 |
| Fluence at 1.3 GHz, $F$ (Jy ms) | 8.0 | 0.7 | 1.8 | 0.6 |
| $S_\nu D^2$ ($\times 10^{12}$ Jy kpc$^2$) | 10.2 | 1.9 | 5.1 | 1.4 |
| Energy Released, $E$ (J) | ~$10^{33}$ | ~$10^{31}$ | ~$10^{32}$ | ~$10^{31}$ |

**Table 1. Parameters for the four FRBs.** The position given is the center of the gain pattern of the beam in which the FRB was detected (half-power beam-width ~ 14 arc min). The UTC corresponds to the arrival time at 1581.804688 MHz. The DM uncertainties depend not only on SNR but also on whether $\alpha$ and $\beta$ are assumed ($\alpha = -2$; no scattering), or fit for; where fitted, $\alpha$ and $\beta$ are given. The co-moving distance was calculated using $DM_{Host} = 100$ cm$^{-3}$ pc (in the rest frame of the host) and a standard, flat-universe $\Lambda$CDM cosmology which describes the expansion of the universe with baryonic and dark matter, and dark energy ($H_0 = 71$ km s$^{-1}$ Mpc$^{-1}$, $\Omega_M = 0.27$, $\Omega_\Lambda = 0.73$; (*29*) $H_0$ is the Hubble constant, and $\Omega_M$ and $\Omega_\Lambda$ are fractions of the critical density of matter and dark energy respectively). $\alpha$ and $\beta$ are from a series of fits using intrinsic pulse widths of 0.87 to 3.5 ms; the uncertainties reflect the spread of values obtained (*2*). The observed widths are shown; FRBs 110627, 110703, and 120127 are limited by the temporal resolution due to dispersion smearing. The energy released is calculated for the observing band in the rest frame of the source (*2*).

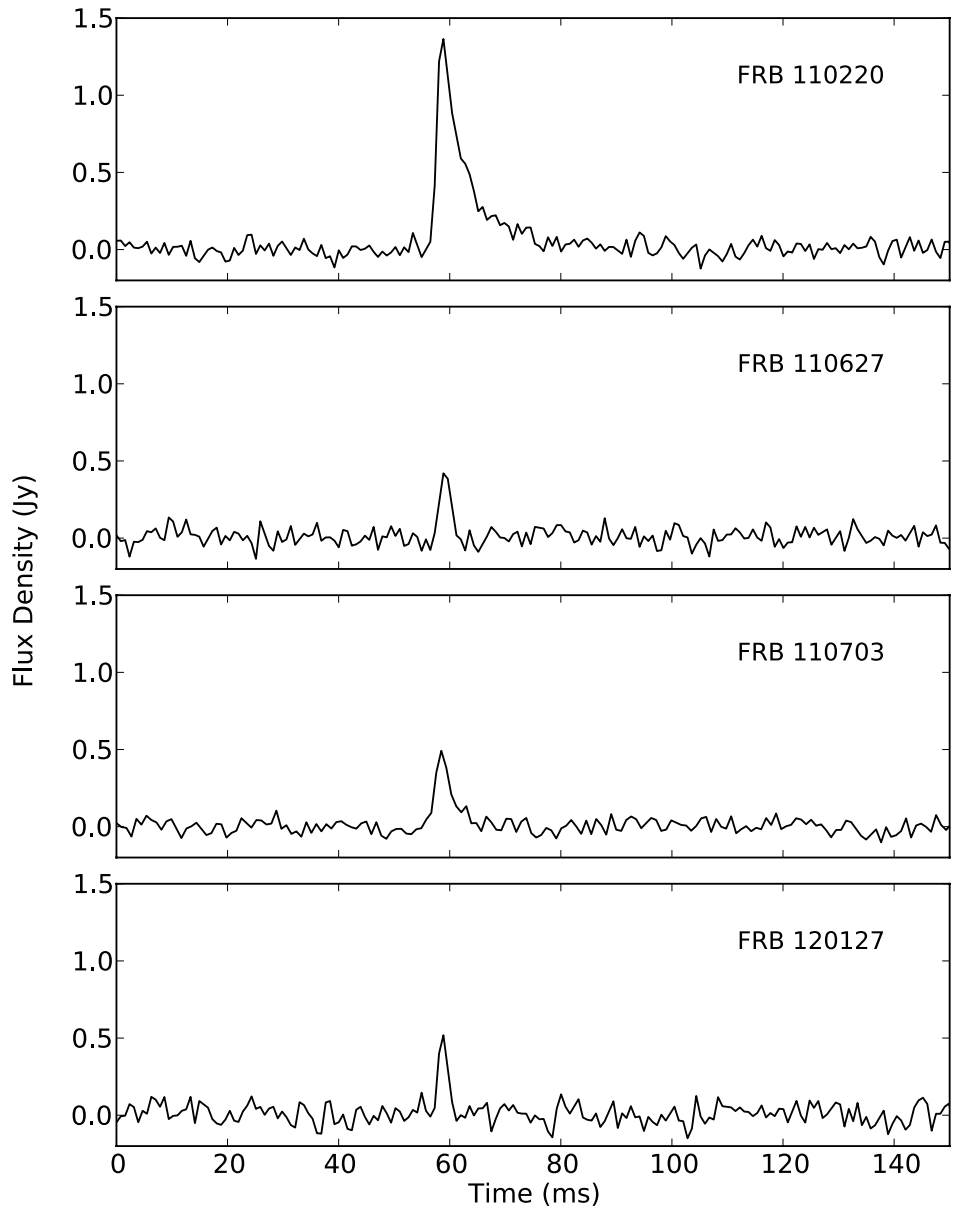

**Fig. 1. The frequency-integrated flux densities for the four FRBs.** The time resolutions match the level of dispersive smearing in the central frequency channel (0.8, 0.6, 0.9, and 0.5 milliseconds, respectively).

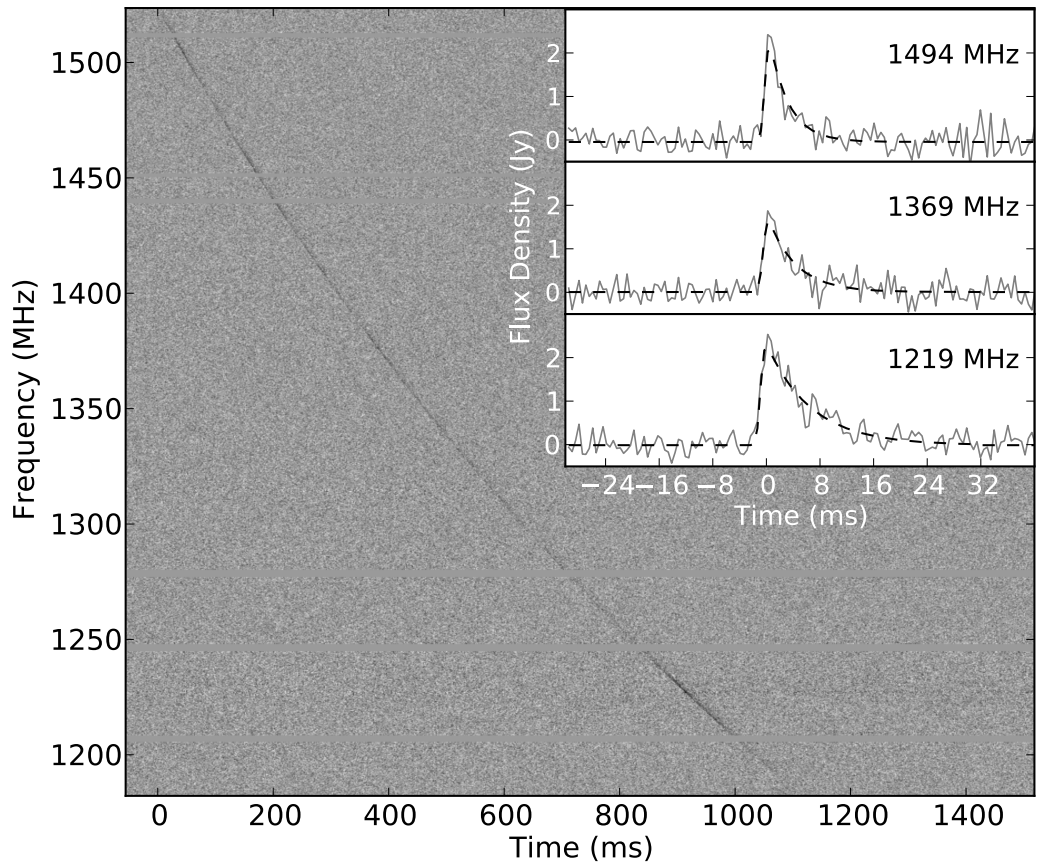

**Fig. 2. A dynamic spectrum showing the frequency-dependent delay of FRB 110220.**
Time is measured relative to the time of arrival in the highest frequency channel. For clarity we have integrated 30 time samples, corresponding to the dispersion smearing in the lowest frequency channel. (**Inset**) The top, middle, and bottom 25-MHz-wide dedispersed sub-band used in the pulse-fitting analysis (*2*); the peaks of the pulses are aligned to time = 0. The data are shown as solid grey lines and the best-fit profiles by dashed black lines.

**Materials and Methods**

The High Time Resolution Universe (HTRU) survey at Parkes is separated into high-, mid- and low-Galactic latitudes (*1*). These Fast Radio Bursts (FRBs) have been discovered in the high-latitude region that was designed to provide a snapshot of the transient radio sky (see Table S1).

The survey uses the 13-beam receiver at the prime focus of the 64-m Parkes radio telescope, New South Wales, Australia. Each beam has a half-power beam-width (HPBW) at 1.3 GHz of approximately 14 arcminutes. Adjacent beam centers are separated by approximately 2 HPBW and as such a single pointing of the telescope does not cover a patch of sky at the HPBW gain level. Consequently, when an FRB is detected in a single beam the closest simultaneously observed beam centers are 2 HPBWs away. Multiple telescope pointings are required to fully sample the sky (*1*).

The mid-latitude part of the survey has been completed, while observations for the low- and high-latitude regions are ongoing at the time of writing. The search for single pulses in the high-latitude region has covered 95,670 individual beams each observed for 270 seconds. This corresponds to 24% of the survey region and almost 10% of the full sky; it is in this data set that the FRBs were found. The processed high-latitude beams are concentrated in the region $-70° < b < -30°$ corresponding to lines of sight with low Milky Way (MW) contributions to dispersion measure (DM) and scattering.

The digital back-end 8-bit samples two polarizations from each beam, the sum of these two values is then 2-bit sampled. This is done across 1024 frequency channels every 64 μs. The data are recorded to tape and sent to Jodrell Bank Centre for Astrophysics and Swinburne University of Technology for processing and searching. The data first undergo a process of radio frequency interference (RFI) excision; the first stage of which is to flag channels containing bright, repetitive signals. Secondly, the time-series corresponding to the average of all channels, without any correction for dispersion applied, is searched for bright bursts; any found are removed from the data; in this way non-dispersed terrestrial bursts are rejected. There are channels that always contain RFI at Parkes; these are flagged automatically. Typically around 190 channels are flagged.

The data are then "dedispersed" by correcting for delays corresponding to approximately 1400 DM values ($0 < DM < 2000$ $cm^{-3}$ pc) before averaging all channels. The resultant one-dimensional time-series are then searched for significant peaks with a series of matched filters of increasing width; these filters improve sensitivity to bursts wider than the sampling resolution. Signals with signal-to-noise ratio (SNR) < 6 are disregarded to reduce the number of spurious candidates due to thermal noise (*20*). In this way candidate single pulses are generated. The HTRU survey is less sensitive to high-DM single pulses because of the dispersive spread within one channel is increased, this cannot be corrected for (*20*).

At this stage the candidates resulting from the 13 simultaneous beams of the receiver undergo a multibeam rejection routine. This involves removing candidate pulses appearing in more than 9 beams; this is done in an effort to reject RFI, and in particular perytons, which appear in all beams (*8*). A final cut (DM > 100 cm$^{-3}$ pc and SNR > 9) is applied before human inspection of all remaining candidates.

Pulse Fitting Method

The full bandwidth was divided into a number of sub-bands. Each sub-band was convolved with a template with a scattering tail using a characteristic scattering time. This scattering time was determined from a value at a reference frequency, 1 GHz, which was scaled according to $\tau \propto \nu^{\beta}$ to each sub-band center frequency. Note $\tau$ is the broadening due to scattering, and is different from the measured width, *W*, which may include intrinsic and other broadening effects. Using the arrival time of the burst at reference frequency $\nu_0$, the arrival time at a frequency $\nu$ was scaled according to a dispersion law $\delta t \propto \mathrm{DM} \cdot \nu^{\alpha}$. The parameters $\delta t(\nu_0), \tau(\nu_0)$, DM, $\alpha$, and $\beta$, were determined in a least-square fit using the SIMPLEX and MIGRAD algorithms from CERN's MINUIT package (*30*). Uncertainties were derived using the MINUIT algorithm to explore the error matrix, which also attempts to account for correlations between parameters.

An overall baseline and amplitude of the scattered pulse of each sub-band were also treated as free parameters to be fit. Before the final step, these parameters were kept fixed to the previous best-fit value.

**Supplementary Text**

Fitting Results

FRB 110220 has the largest SNR and allows the determination of all parameters as described above. Here, the band was divided into 16 sub-bands, two sub-bands at the top and one at the bottom of the frequency range were excluded from the fit, as, due to an instrumental roll-over of the bandpass, SNR was low. The resulting 13 sub-bands, with 25 MHz each, are frequency-averaged versions of the original 0.39 MHz-wide channels in each sub-band. The time resolution was rebinned to 0.512 ms per sample. The fitting results are consistent with an unresolved pulse at infinite frequency, so a Gaussian pulse was used as a template. The width of the pulse was varied according to the dispersion smearing at each sub-band, ranging from 0.87 ms (1.7 samples) to 1.80 ms (3.5 samples). To take this variation of the width into account before convolving with the scattering tail, the fitting was performed for various widths in this range. In the case of $\alpha$ these fits indicated a small spread of values about that given in Table 1, the uncertainty specified is the spread of these values; the statistical uncertainty on each fit was significantly smaller, as indicated in Figure S1. For $\beta$ the spread of values was of similar magnitude to the statistical uncertainty on a single fit; in this case the uncertainty of the value in Table 1 reflect the spread of $\beta$ and the statistical uncertainty on a single fit which are of a similar magnitude.

Alternatively, the template could have been spread by dispersion within each sub-band, but as the dispersion law was included in the fit, and as the true intrinsic shape of the pulse is not known, we used variation of the template width as a means to explore the dependence of our results on template width.

All of the intrinsic pulse width fits indicated a similar $\chi^2$ including unresolved (limited by the downsampled temporal resolution). The marginally best $\chi^2$ was obtained for a dispersed (within the sub-bands) template width of 2 bins (1.02 ms), i.e $\chi^2$ = 26316 for 26605 degrees of freedom, corresponding to a reduced $\chi^2$ of 0.989, indicating an excellent fit (see Fig. S1). Taking the dispersion smearing into account, this suggests an intrinsic width of less than 1.02 ms. Overall, the fit is consistent with the expectation from a cold plasma law and interstellar scattering, confirming the astrophysical nature of FRB 110220 (see Fig. S1).

FRBs 110627 and 120127 are too weak for a detailed analysis. FRB 110703 was weaker than FRB 110220 and had to be frequency collapsed to just eight sub-bands of 50 MHz bandwidth, 2 sub-bands were excluded at the top and bottom of the band. The rebinning factor in time was also larger than for FRB 110220 in order to achieve sufficient SNR. We found, for the resulting 3.512 ms resolution, the pulse was consistent with not having been scattered. Attempts to increase the time resolution in order to resolve the pulse resulted in an SNR too low to obtain reliable fit results. We therefore only adapted one template width, namely 0.57 bins (2.01 ms), which corresponds to the DM smearing at the lowest frequency channel, and fitted in the final step only for $\delta t(v_0)$, DM, and $\alpha$. Again, the results support the astrophysical nature of the pulse. Note in Table 1 the uncertainty on $\alpha$ is taken to be the uncertainty for FRB 110220. We do this because we have not fit for $\beta$ simply because there is insufficient SNR; there is likely some covariance between $\alpha$ and $\beta$ which is implicit in the value derived for $\alpha$.

RFI

It is important and difficult to distinguish non-repeating radio sources from RFI to be sure of an astronomical origin. Knowledge of well-studied astronomical signals of similar type, or which have undergone similar propagation effects, inform us about what to expect from such a signal. While FRBs are not from pulsars they share the characteristics of broadband coherent emission that has propagated through turbulent ionized material. The effects of this propagation on pulses emitted by pulsars are well studied, and the common observables are well understood. The FRBs are measured to obey astronomical dispersive delay and scattering to a high accuracy; it seems unlikely that a man-made source, especially emitting in a protected band, would so accurately reproduce this behavior. The broadband nature of the bursts also rules out the majority of the RFI environment, which consists of short-duration non-dispersed narrowband pulses.

The ~25 now-reported perytons (*8, 19, 31*) are characterized by 20–50 ms wide, symmetric, swept-frequency pulses that imperfectly mimic a $\delta t \propto v^{-2}$ dispersive sweep. Perytons' equivalent DM distribution has a peak near 375 $cm^{-3}$ pc and a negative skew (the range is ~ 200–420 $cm^{-3}$ pc). Perytons are easily recognized by their appearance in all beams of the multibeam receiver, indicating a sidelobe detection. The similarity of these peryton properties with the sweep rate and duration of the FRB 010724 led to the suggestion it might be a peryton during which the telescope was directly pointed at the source, however its celestial or terrestrial origin cannot yet be conclusively determined (*8*). This is not the case for the four FRBs presented here, which have a much larger (and apparently random) DM range, adhere to a dispersive sweep to high precision, and three of which are a factor of > 3 shorter duration than all known perytons. Furthermore, the fast-rise and exponential-tail profile of FRB 110220 makes a clear case for cold plasma propagation; FRB 010724, while displaying frequency-dependent pulse broadening, did not have clear asymmetry.

Because spatial filters are powerful discriminators of local RFI, the current search system rejects signals that are significantly detected in nine or more of the telescope's 13 beams. Therefore, we are likely filtering out a large number of perytons expected to appear in the HTRU survey. Thus far, two perytons (sufficiently weak to not be detected in all beams due to receiver gain variations) have been discovered in HTRU data from the year 2010. This indicates that perytons are still occurring at the telescope, however it also confirms that perytons still exhibit the same imperfectly-dispersed, symmetric form, with <DM> ~ 375 cm$^{-3}$ pc and <W> ~ 30 ms. Like all previously-detected perytons, the new perytons are not obvious analogs to FRBs.

DM contributions

Signals that originate in the MW, and are dispersed by the interstellar medium (ISM), are not redshifted. In this case, $\mathrm{DM} = \int n_e \cdot dl$ assumes $dl$ is constant during signal propagation. The MW delay contributions are therefore constant with source redshift for a given line of sight because the frequencies sampled at the telescope are essentially the same as they were when they reached the edge of the MW. The DM contribution from the MW is highly dependent on Galactic latitude, at high Galactic latitudes there is little material along the line of sight and typical MW DMs are small. FRBs at high Galactic latitudes exhibit extremely high DMs that are clearly anomalous to the Galactic pulsar population: a distinct indication of their extragalactic nature (see Fig. S2).

If the sources are located in a galactic disk, and the MW model of the free electron distribution (*10*) is representative of the host galaxy, then $\mathrm{DM}_{\mathrm{Host}}$ is dictated by the viewing angle. Considering a median inclination angle, $i = 60°$, the galaxy's ISM causes $\mathrm{DM}_{\mathrm{Host}} \approx 100 \mathrm{~cm}^{-3}$ pc ; therefore, the majority of $\mathrm{DM}_E$ would be due to the IGM.

It may also be possible that an intervening galaxy along the line of sight to the source is an additional source of dispersion. As the measured dispersion is an integrated effect the presence of an intervening galaxy would reduce the contribution to the total DM from the IGM. This intervening galaxy contribution would be subject to the same conditions regarding inclination as described for sources located in a galactic disk. If there were a contribution from an intervening galaxy, this would, of course, equate to a closer source. We note the probability of an intervening source within $z < 1$ is less than 0.05 (*16*).

When considering dispersion caused by the IGM and a host it becomes necessary to take into account the expansion of the universe and the fact that the radiation that makes up the pulse is affected by the stretching of space-time. An important manifestation is that the radiation that underwent dispersion in a host galaxy was at a higher frequency, $\nu \rightarrow \nu(1+z)$. Since the dispersive delays follow $\delta t \propto \nu^{-2}$, the *observed* delay due to a given $\mathrm{DM}_{\mathrm{Host}}$ decreases with increasing redshift. The same argument applies to dispersive delays from the IGM where the electron density and length-scale both depend on redshift. The total delay due to the IGM is found by integrating along the line of sight to a source. By combining this with host and MW delay contributions a total delay can be found and a redshift calculated (see Fig. S3).

FRB locations

Detailed attempts to constrain the position of FRB 110220 by using its non-detection in the other beams does not produce a better localization than the HPBW and is not simply described by a single quantity. Positioning is further complicated because the beam shape is frequency dependent – it is wider for lower frequencies. This imparts an apparent spectral index to a source located away from the beam center. Combining this with the unknown intrinsic spectral index there is little constraint on the location within the beam. For these reasons the HPBW is used while recognizing the difficulty in knowing the true position.

There are 15, 56, 30, and 54 cataloged galaxies within the HPBW for FRB 110220, 110627, 110703, and 120127 respectively. Some of these have measured redshifts of $z = 0.02, 0.6$. If located in one of these galaxies then $DM_{IGM}$ would be either 17 $cm^{-3}$ pc ($z = 0.02$) or 570 $cm^{-3}$ pc ($z = 0.6$), as such a significant fraction of $DM_E$ would be due to the host. In these directions there are almost certainly as yet undiscovered galaxies at other redshifts.

The empirical relationship between DM and scattering (*15*) may be used to constrain the redshift to $z > 0.11$

In addition to looking at galaxies in the field of view we also looked for temporally associated GRBs or x-ray transients. Gamma-ray and x-ray telescopes have all-sky monitors however we did not find any cataloged event (NASA HEASARC) which could be related to any of the FRBs in this paper. The Astronomers Telegram was also checked for possibly associated events; none were found.

Energetics

The energy released, E, (see Table 1) is calculated as the total energy of the burst at the source. The energy released, $E \sim F \cdot D^2 \cdot BW$, where $F$ is the fluence, $D$ is the cosmological luminosity distance, and $BW$ is the bandwidth at the source. Note that emission is not assumed isotropic, instead a beam of 1 steradian is used for simplicity. In the case of these FRBs, where there is significant redshift, the width, distance, and bandwidth must be corrected to account for cosmological redshift. These corrections have been made for the $E$ values in Table 1. As the measurement of spectral index is inconclusive we have assumed it to be flat (Fig. S4) and do not extrapolate beyond the observing band.

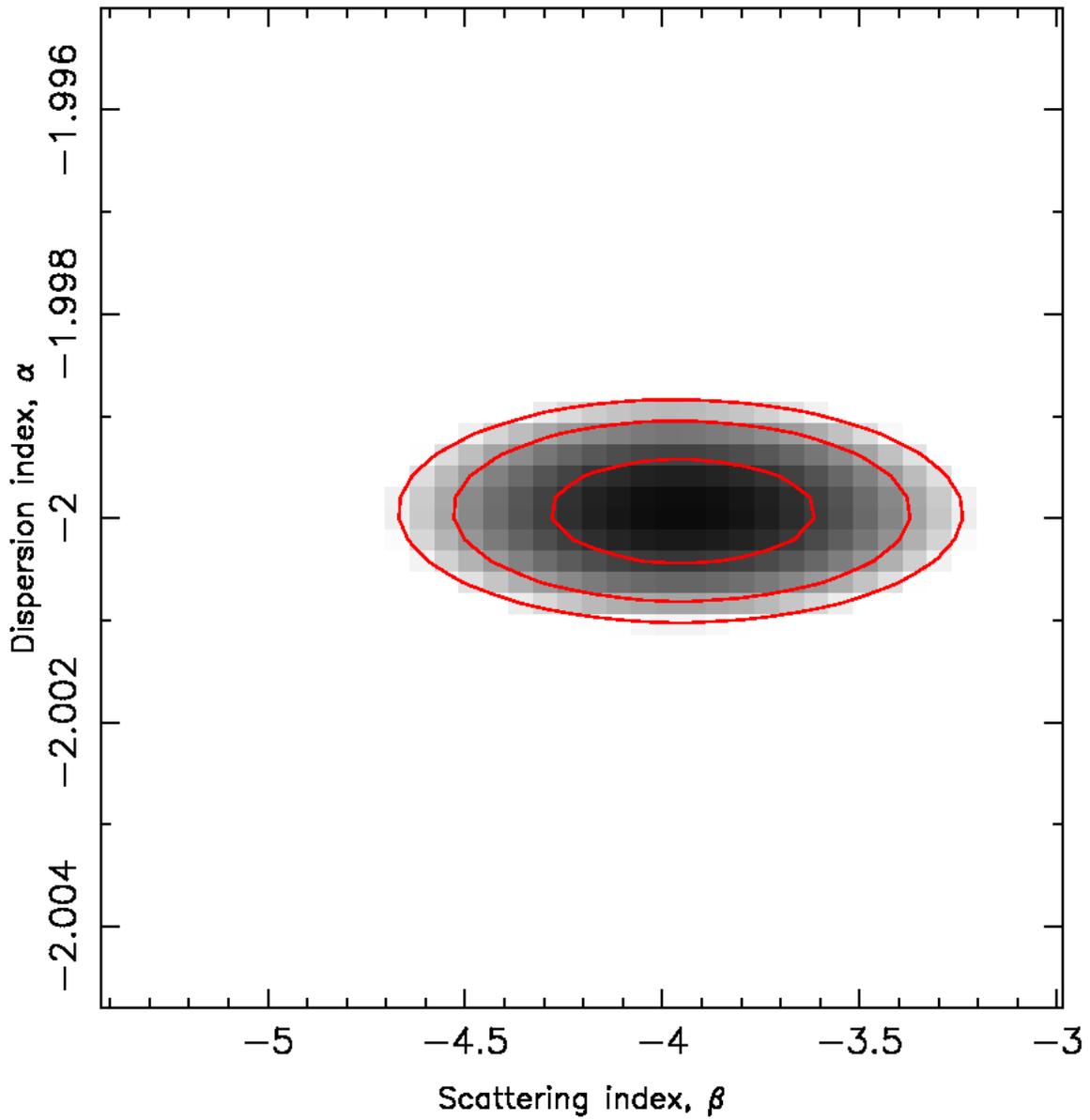

**Fig. S1.**
A grayscale plot showing the reduced-$\chi^2$ as a function of dispersion index, $\alpha$, and scattering index, $\beta$. The concentric lines indicate the 1-, 2-, and 3-$\sigma$ confidence intervals about the best values (see Table 1). Note this figure corresponds to the fit with a smeared intrinsic width of 2 bins, 1.02 ms, not the ranges indicated in Table 1.

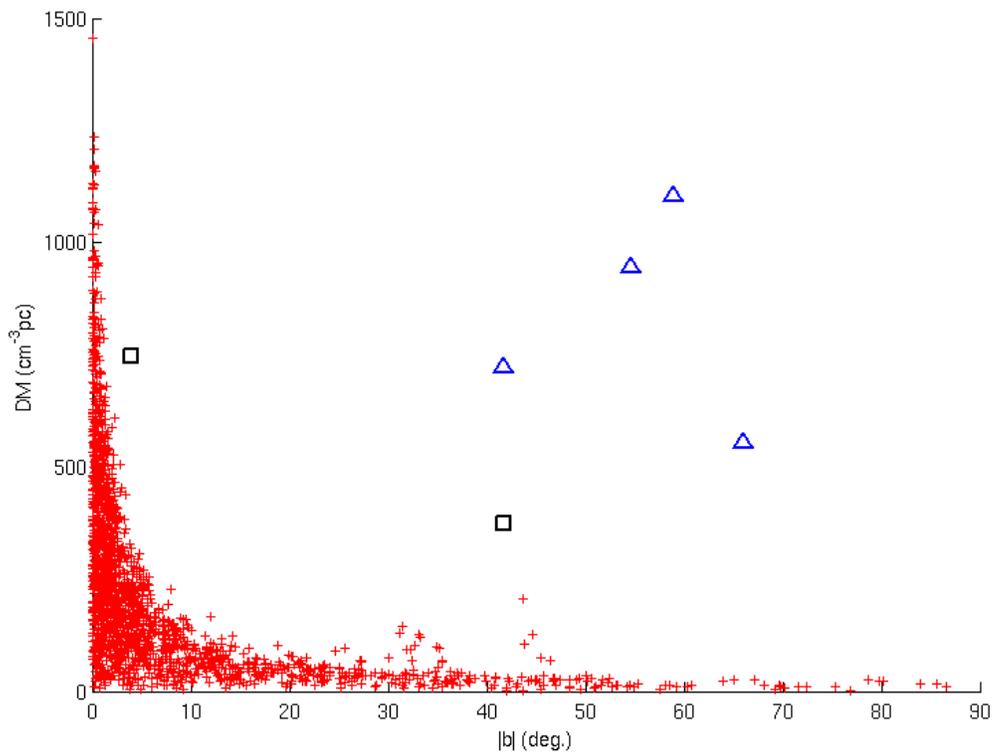

**Fig. S2.**
Measured DM for FRBs and known pulsars is plotted against the magnitude of Galactic latitude. The FRBs from this paper are shown as blue triangles, FRB 010621 and 010724 are shown as black squares, pulsars are indicated by red '+' symbols. The FRBs exhibit significantly higher dispersion than pulsars at similar separations from the Galactic plane. The pulsars with an apparent dispersion excess located at 30° < |b| < 45° are in the Magellanic clouds which provide an additional source of free electrons and dispersion.

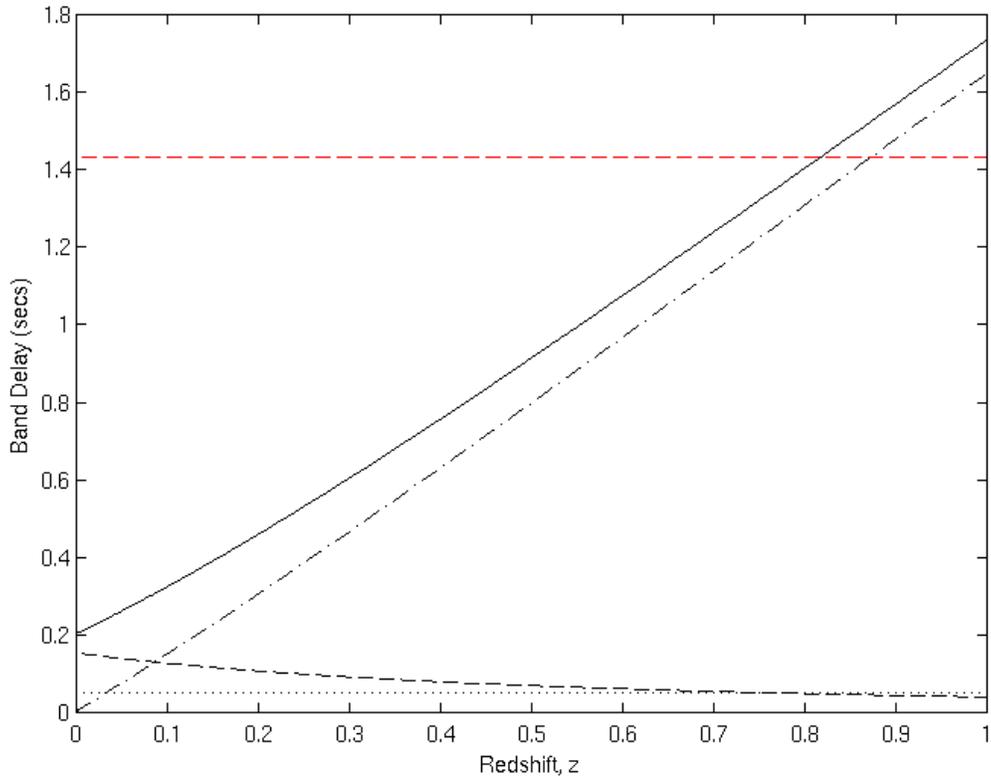

**Fig. S3**
Modelled dispersive delay (in seconds) across the observing band of the HTRU survey is shown plotted against redshift. The flat dashed line indicates the delay across the observing band for FRB 110220. The dotted line is the MW contribution and is constant irrespective of the redshift of the source, the dot-dashed line is the delay due to the IGM, the curved dashed line is the delay contribution from a host galaxy with $DM_{Host} = 100$ $cm^{-3}$ pc, and the solid line is the sum of the IGM, host and MW contributions. When the sum of contributions is equal to the measured delay the redshift is inferred.

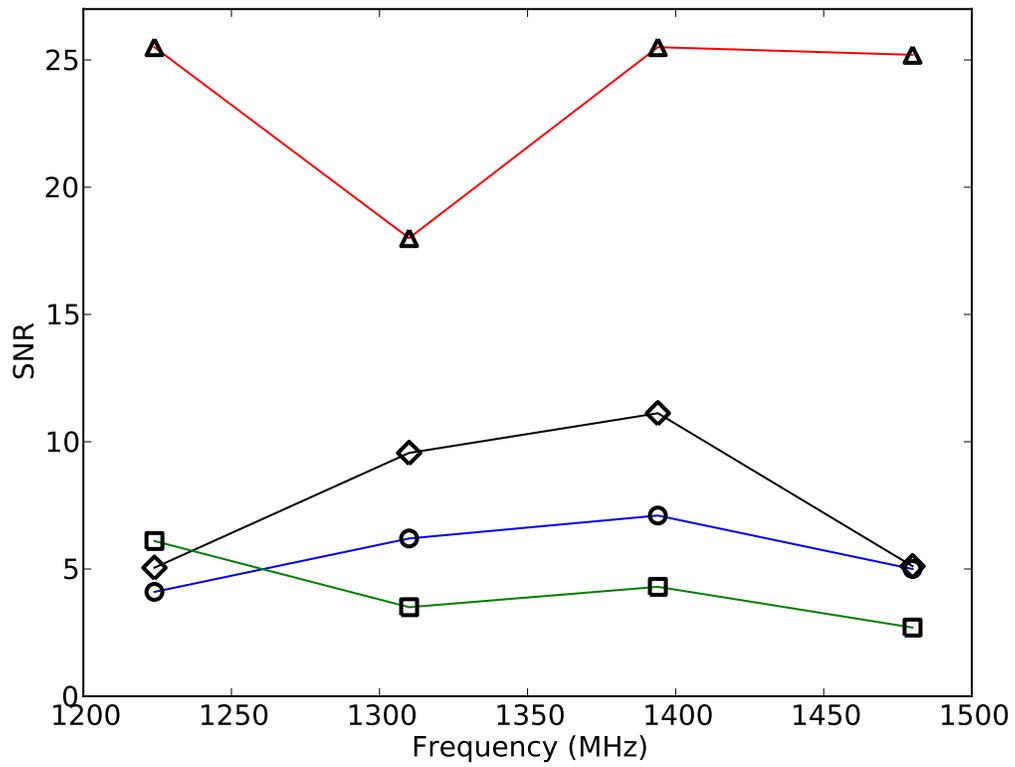

**Fig. S4**
The SNRs of the FRBs (110220: triangles, 110627: circles, 110703: diamonds, 120127: squares). For each the total band has been split into four 100-MHz sub-bands and the SNR of the pulse in each measured.

**Table S1.**

The parameters detailing the HTRU high-latitude southern survey, which covers the sky visible to the Parkes radio telescope. The areas quoted correspond to the total area observed and processed at the time of writing; the area of a single beam, 0.08 deg$^2$, is calculated using the half-power beam-width at 1.3 GHz.

| Region | High-Latitude |
|---|---|
| Coverage | Declination < +10° |
| Pointing Length (s) | 270 |
| Sampling time (μs) | 64 |
| Band Center Frequency (MHz) | 1382 |
| Channel Bandwidth (MHz) | 0.390625 |
| Number of channels | 1024 |
| Beams at completion | 443287 |
| Beams Processed | 95670 |
| FRBs Found | 4 |
| Completed Area (deg.$^2$) | 4078 |
| Completed Time (days) | 298 |
| Completed Area (skys, 4π sr) | 0.098 |
| Completed Time (years) | 0.82 |